\documentclass[11pt]{article}
\usepackage{moriond,epsfig}

\def\beq{\begin{equation}} 
\def\eeq{\end{equation}} 
\def\bea{\begin{eqnarray}}
\def\eea{\end{eqnarray}}

\def\bar#1{\overline{#1}}

\def\inv{^{\raise.15ex\hbox{${\scriptscriptstyle -}$}\kern-.05em 1}} 
\def\lbar{{\lower.35ex\hbox{$\mathchar'26$}\mkern-10mu\lambda}} 

\let\<=\langle 
\let\>=\rangle

\let\+=\uparrow 
 
\newcommand{\newc}{\newcommand} 
\newc{\gsim}{\lower.7ex\hbox{$\;\stackrel{\textstyle>}{\sim}\;$}} 
\newc{\lsim}{\lower.7ex\hbox{$\;\stackrel{\textstyle<}{\sim}\;$}} 
\newc{\gev}{\,{\rm GeV}} 
\newc{\mev}{\,{\rm MeV}} 
\newc{\ev}{\,{\rm eV}} 
\newc{\kev}{\,{\rm keV}} 
\newc{\tev}{\,{\rm TeV}}

\newc{\mz}{M_Z} 
\newc{\mw}{m_{\rm weak}} 
\newc{\nr}[1]{N^c_R{}_{#1}} 
 
\begin{document}
\vspace*{4cm}
\title{ASYMMETRIC SNEUTRINO DARK MATTER}

\author{Stephen M. West}

\address{Rudolf Peierls Centre for Theoretical Physics, University of Oxford,
1 Keble Rd., Oxford OX1 3NP, UK}

\maketitle\abstracts{
It is known that the cosmological baryon density ($\Omega_{\rm{b}}$) and dark matter  
density ($\Omega_{\rm{dm}}$) have strikingly similar values. However, in most theories of the early Universe, each density is explained 
by separate dynamics and consequently there is no compelling reason for this observation. In this note, I briefly review a model in 
which the dark matter possesses a particle-antiparticle asymmetry. This asymmetry determines both the baryon asymmetry and strongly 
affects the dark matter density, thus naturally linking $\Omega_{\rm{b}}$ and $\Omega_{\rm{dm}}$. 
In these models it is shown that sneutrinos can play the role of such dark matter.}

\section{Introduction}
For some time it has been apparent that the inferred values of the 
cosmological baryon and dark matter densities are strikingly similar. 
The WMAP-determined range \footnote{The analyses presented in this note does not include the most recent WMAP data \cite{wmap3}.} for 
the dark matter density, 
$0.129 > \Omega_{\rm{dm}} h^2 > 0.095$, is within a factor of a few 
of the combined WMAP and big-bang nucleosynthesis determined value 
of the baryon density \cite{WMAP,BBN}, $0.025 > \Omega_{\rm{b}} h^2 > 0.012$.

In the vast majority of models of the early universe, the cosmological 
baryon and dark matter densities are independently determined. 
The surviving baryon density is set by a baryon asymmetry generated  
during baryogenesis, and thus depends upon unknown baryon-number violating dynamics and unknown CP-violating phases. In contrast, the 
dark matter density is set by the `freeze-out' 
of the interactions that keep the dark matter in equilibrium, and 
is independent of the dynamics of baryogenesis. Consequently, there is no reason why we should expect $\Omega_{\rm{b}}$ and 
$\Omega_{\rm{dm}}$ to coincide.   

One possible solution to this problem is to link
the dynamics of baryogenesis with that of the origin of dark matter. 
In particular, it is natural to consider models where the dark matter 
and baryon sectors share a quantum number, either continuous or discrete, 
which provides a relation between their surviving number densities and thus 
energy densities.

Specifically, in \cite{hmw}, we proposed models of dark matter 
possessing a particle-antiparticle asymmetry, where this 
asymmetry strongly affects the dark matter density, and 
through the electroweak (EW) anomaly, determines the baryon asymmetry, thus
naturally linking $\Omega_{\rm b}$ and $\Omega_{\rm dm}$\footnote{For an early attempt along these lines see~\cite{kaplan}.}. 

In this model, assuming the particle-particle annihilation cross section is negligible, we are able to write down a simple relationship 
between $\Omega_{\rm{b}}h^2$ and $\Omega_{\rm{dm}}h^2$ given by \cite{hmw},
\begin{equation} 
\Omega_{\rm{dm}} h^2 = \Omega_{\rm{b}} h^2 
\frac{A}{A_{\rm{bary}}} \frac{m}{m_{\rm{bary}}}, 
\label{estimate} 
\end{equation} 
where $A$ and $A_{\rm{bary}}$ are the particle-antiparticle 
asymmetries of the proposed dark matter relic and of baryons, defined by $A=(n - \bar{n}) 
/n$. Here $m$ and $m_{\rm{bary}}$ are the masses of our dark matter relic and of baryons ({\it i.e.} 
the proton mass). The ratio of $A$ to $A_{\rm{bary}}$ is determined by the "chemical" equilibrium conditions between the two sectors 
just before the freeze-out of the relevant interactions. 

If the particle-particle annihilation cross section for the relic is not negligible, Eq.(\ref{estimate}) will 
not hold, although there will be a generic tendency for the density of the relic 
to move towards this value as a result of an asymmetry. For full details of how a matter-antimatter asymmetry affects the  
density of a thermal relic see \cite{hmw} and references therein.

\section{The Model: Mixed Sneutrino Dark Matter}

In \cite{hmw} it was shown that sneutrinos can play the role 
of such dark matter in a previously studied variant 
of the MSSM. In this model the light neutrino masses result from 
higher-dimensional supersymmetry-breaking terms
\cite{ahhmsw,borz,sw,mrw}.  This model preserves all the successes of the MSSM, such  
as stability of the weak scale and unification of gauge couplings,  
while being, at least in part, testable at the LHC. 

Within the context of the Minimal Supersymmetric Standard Model 
(MSSM), sneutrinos do not make a very appealing dark matter 
candidate. Sneutrinos tend to annihilate too efficiently, 
resulting in a relic density smaller than the observed dark matter 
density. Furthermore, their elastic scattering cross section is sufficiently large to be easily 
observed by direct dark matter experiments. 

In the models of \cite{ahhmsw,borz,sw,mrw} the left-handed `active' sneutrino, $\tilde{\nu}$,
mixes, via large $A$-terms, with the right-handed `sterile' sneutrino state, $\tilde{n}$,
producing the light mass eigenstate given by $\tilde{\nu}_1=-\tilde{\nu}\sin\theta 
+ \tilde{n}^* \cos\theta$, where $\theta$ is a mixing angle. This mixing reduces the annihilation cross section, potentially 
providing the appropriate quantity of dark matter. In addition, since the coupling of the lighter sneutrino eigenstate, 
$\tilde{\nu}_1$, to the $Z$ is suppressed by $\sin \theta$, the direct LEP experimental 
constraints are weakened.  


Another important feature of these models is that 
the light sneutrino states share a non-anomalous $(B-L)$-symmetry 
with the baryons which is only weakly broken by the Majorana 
neutrino masses. It is this approximately conserved symmetry which provides the link between the dark matter and baryon number 
densities. 
  
Turning to the calculation of the relative asymmetry in the 
sneutrino and baryon sectors, the method is a simple adaptation 
of the standard "chemical" equilibriation techniques applied 
to, for example, the calculation of the ratio $B/(B-L)$ in the 
MSSM~\cite{iims} in the presence of
anomaly-induced baryon number violating processes 
in the early universe. 

In this analysis we assume that at a temperature $T$ (with 
$T>T_c$, where $T_c$ is the electroweak phase transition temperature) the MSSM susy spectrum, including $k$ 
rhd sneutrinos can be considered light ($m\lsim T$). 

The resulting relative asymmetry in the 
sneutrino and baryon sectors is given by, \cite{hmw}
\beq 
\frac{A}{A_{\rm bary}} = \frac{k}{3}~~{\rm to} ~~\frac{k}{6},
\label{asymresult} 
\eeq 
where the variation depends upon the spectrum of sneutrino masses with respect to $T_c$.  In what follows we specialize to the case in 
which $k=1$.

In the end the vital point is that it does not matter what the dynamics are which generate the asymmetry at scales $E>T_c$ or indeed 
whether the asymmetry is generated in the baryon or neutrino or sneutrino sector. The $(B+L)$-anomaly-induced interactions 
together with EW gaugino and $A$-term interactions 
automatically distribute the asymmetry between the baryons and 
the light sneutrino states, with a predictable $A/A_{\rm{bary}}$ 
ratio.  
 
The asymmetry could originate from a GUT-based baryogenesis mechanism, or maybe 
more interestingly in the context of the sneutrino dark matter 
model there is the possibility that it could arise via a 
resonant leptogenesis mechanism at the TeV-scale as 
discussed in Ref.\cite{hmrw}.  The end result of the (B+L) violating "chemical" equilibriation process is that 
we expect $1/3 \gsim A/A_{\rm{bary}} \gsim 1/6$ independent 
of the source of the asymmetry.

\section{Results and Discussion}
\begin{figure}[tb] 
\centering\leavevmode 
$\begin{array}{c@{\hspace{0.5in}}c} 
\includegraphics[width=0.28\textwidth]{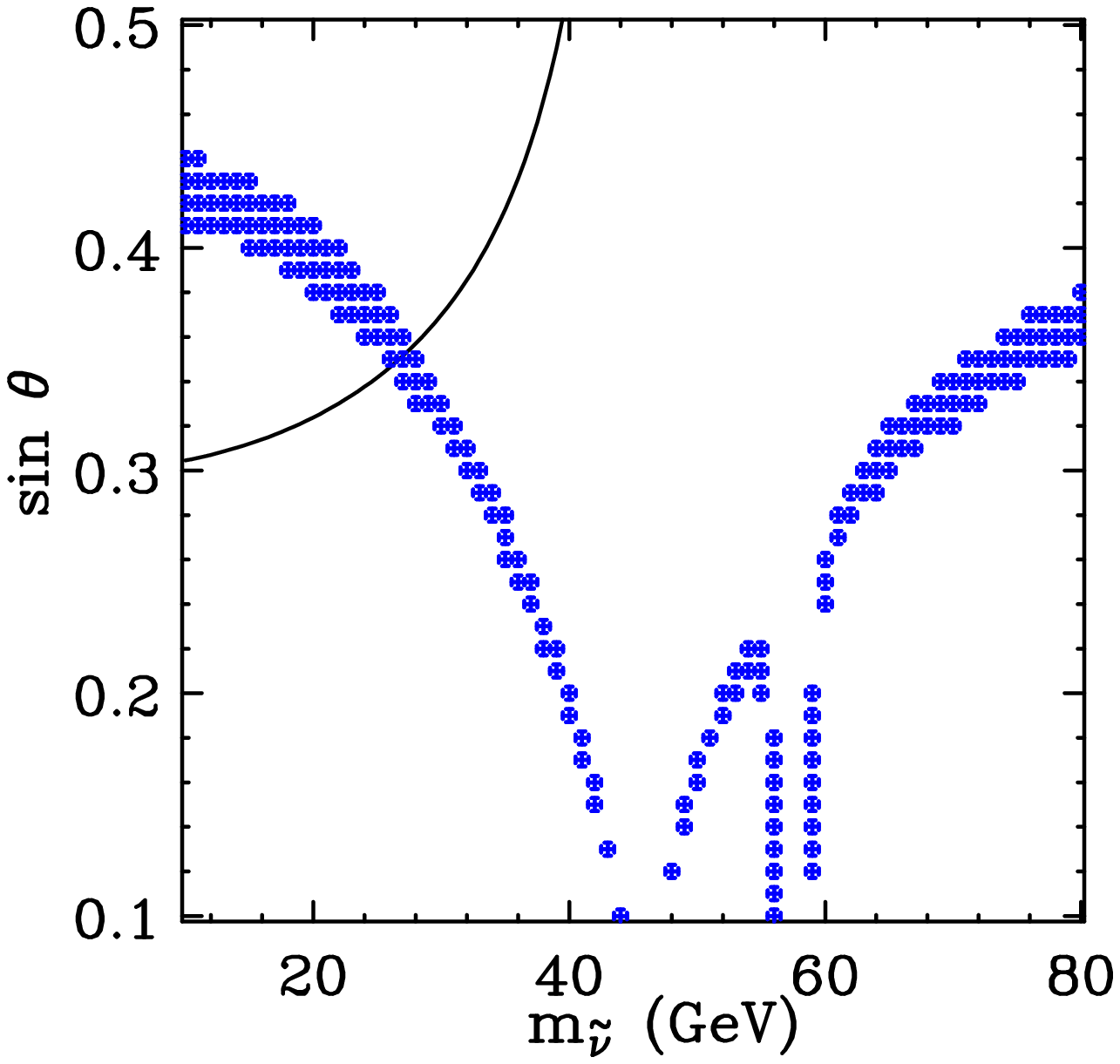}  
\includegraphics[width=0.28\textwidth]{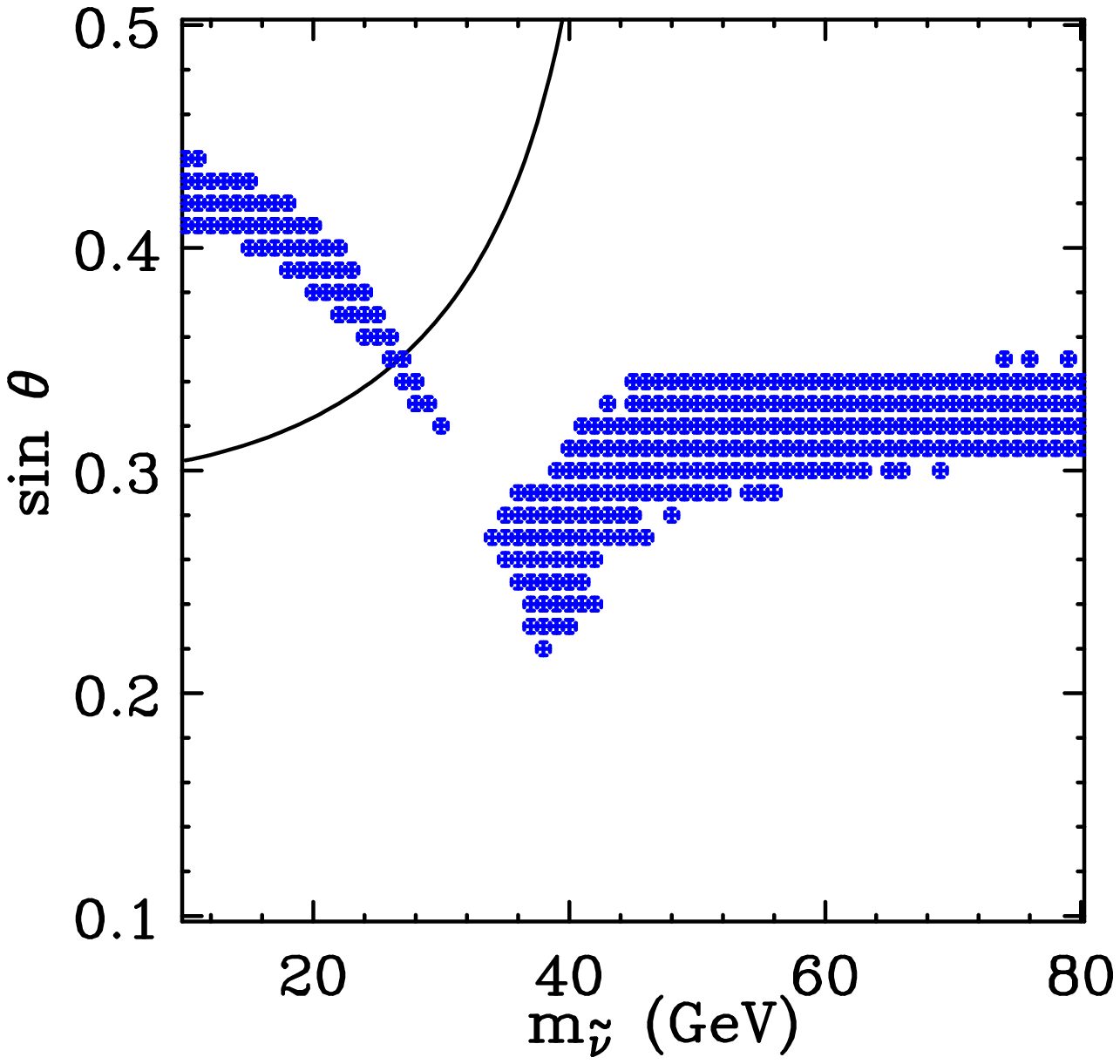}  
\includegraphics[width=0.28\textwidth]{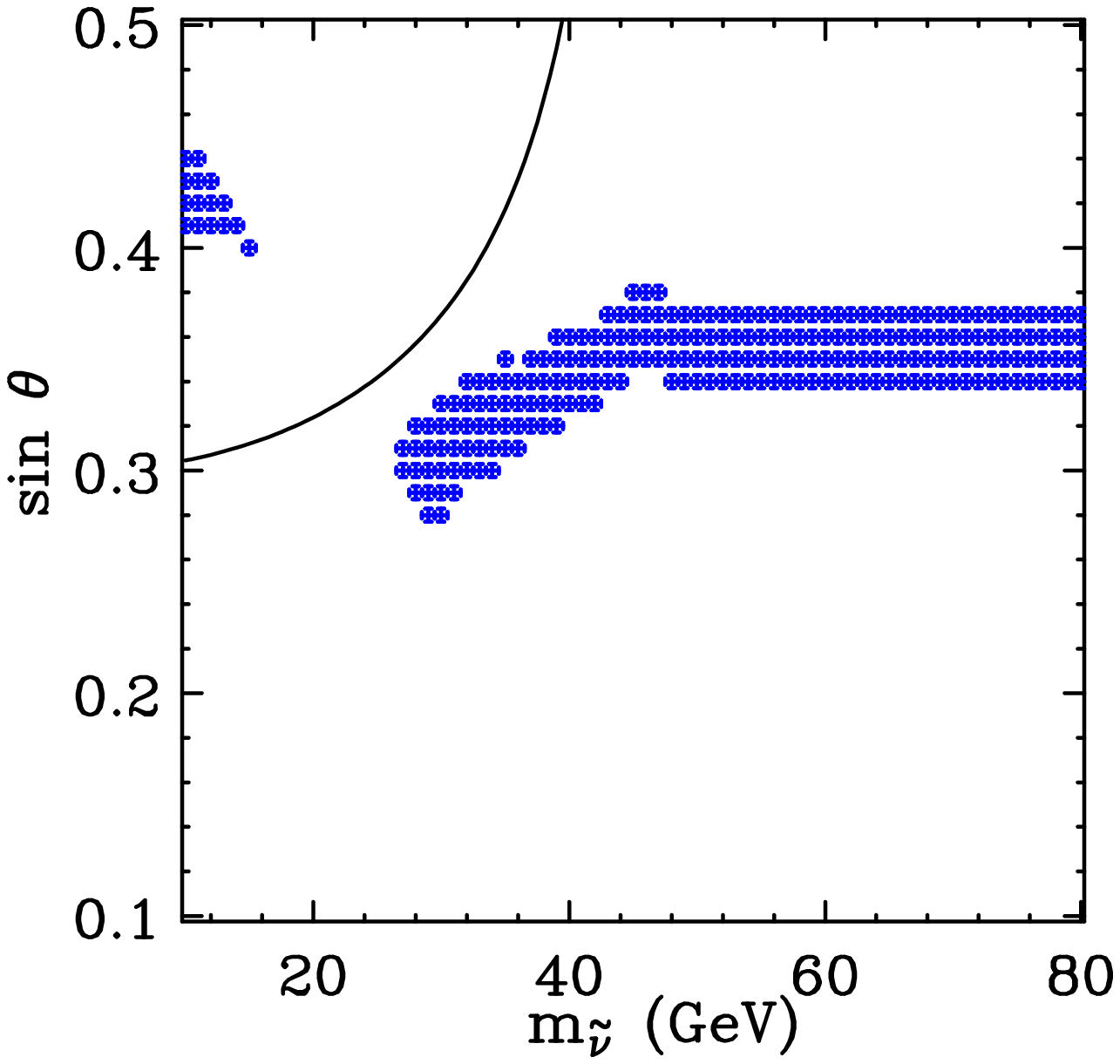}\\ [0.4cm] 
\end{array}$ 
\caption{Parameter space which provides the quantity of 
mixed sneutrino cold dark matter measured by WMAP, $0.129 > 
\Omega_{\rm{dm}} h^2 > 0.095$. In the left frame, the standard 
calculation with no matter-antimatter asymmetry is used. In the center and right 
frames, a dark-matter matter-antimatter asymmetry with 
$A/A_{\rm{bary}} \simeq 1/6$ and $A/A_{\rm{bary}} \simeq 1/3$ respectively is included.
In all cases the measured baryon asymmetry ($\Omega_{\rm{b}}$) is used as an imput. Thus in the shaded regions the observed 
$\Omega_{\rm{b}}/\Omega_{\rm{dm}}$ is reproduced. We use the parameters: $M_1$=300 GeV, $M_2$=300 GeV, 
$\mu$=600 GeV, $\tan \beta=50$ and $m_h$=115 GeV. The region above the 
solid line in each frame is excluded by measurements of the invisible $Z$ decay 
width at LEP} 
\label{relicplot} 
\end{figure}

Our results are shown in figure~\ref{relicplot}. The shaded regions of 
the parameter space predict a relic density within the range measured 
by WMAP ($0.129 > \Omega_{\rm{dm}} h^2 > 0.095$). In the left frame, 
no matter-antimatter asymmetry was included. The dip at 56-59 GeV is due to s-channel higgs exchange to $b 
\bar{b}$. In the center and right frame, a matter-antimatter asymmetry of 
$A/A_{\rm{bary}} \simeq 1/3$ and $A/A_{\rm{bary}} \simeq 1/6$ respectively was included. 

\begin{figure}[tb] 
\centering\leavevmode 
\includegraphics[width=0.30\textwidth]{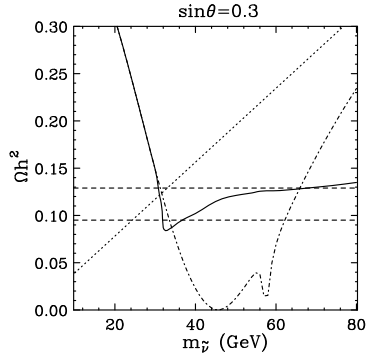}  
\caption{The thermal relic density as a function of mass for 
sneutrinos and anti-sneutrinos with no asymmetry (dot-dash), with a 
matter-antimatter asymmetry of $A/A_{\rm{bary}} \simeq 1/6$ (solid) 
and the estimate of Eq.(\ref{estimate}) (dots). The relic density range 
favored by WMAP is bound by dashed lines ($0.129 > \Omega_{\rm{dm}} 
h^2 > 0.095$). Here we use $\sin \theta$=0.3, $M_1$=300 GeV, $M_2$=300 GeV, 
$\mu$=600 GeV, $\tan \beta=50$ and $m_h$=115 GeV have been used.} 
\label{cutplot} 
\end{figure}

To further illustrate this effect, the result of this calculation across one value of 
$\sin \theta$ is plotted in figure~\ref{cutplot}. Below about 30 GeV, the 
matter-antimatter asymmetry has little effect on the calculation and 
the solid and dot-dashed lines fall nearly on top of each other. In 
the range of roughly 30-70 GeV, however, the asymmetry pulls the relic 
density above the standard symmetric result into the range favored by 
WMAP. Above this range, sneutrino-antisneutrino annihilation 
decreases, leading to larger relic densities for the case with no 
asymmetry. The relic density for the asymmetric case, however, is 
largely determined by the sneutrino-sneutrino annihilation cross 
section in this region, so does not increase as rapidly, therefore
resulting in a relic density much closer to the preferred value, even
for $m_{\tilde{\nu}}>70$~GeV.

\section{Summary}
In the standard freeze-out calculation for a weakly interacting dark  
matter relic, there is little reason to expect a density of dark matter  
which is similar to the density of baryons.  
One possible solution is to introduce an  
asymmetry between dark matter particles and anti-particles which is  
related to the baryon-antibaryon asymmetry. This leads to a natural dark  
matter relic density of the same order of magnitude as the baryon density. 
 
As an example, we considered a mixed sneutrino dark matter candidate which  
transfers its particle-antiparticle asymmetry to the baryons through 
the electroweak  anomaly. The relic density calculation for such a candidate  
has extended and natural regions in the $\sin \theta$ and $m_{\tilde{\nu}}$ parameter space in which the observed 
$\Omega_{\rm{b}}/\Omega_{\rm{dm}}$ is reproduced.
\section*{Acknowledgments}
I would like to thank my collaborators, John March-Russell and Dan Hooper as well as the organisers of the XLIst Rencontres de Moriond 
for an excellent conference and financial support.

\end{document}